\documentstyle[12pt]{article}

\def\fun#1#2{\lower3.6pt\vbox{\baselineskip0pt\lineskip.9pt
        \ialign{$\mathsurround=0pt#1\hfill##\hfil$\crcr#2\crcr\sim\crcr}}}

\renewcommand\({\left(}
\renewcommand\){\right)}
\renewcommand\[{\left[}
\renewcommand\]{\right]}

\newcommand\eq[1]{Eq.~(\ref{#1})}
\newcommand\eqs[2]{Eqs.~(\ref{#1}) and (\ref{#2})}

\newcommand\eqst[2]{Eqs.~(\ref{#1})--(\ref{#2})}

\newcommand\ee{\end{equation}}
\newcommand\be{\begin{equation}}
\newcommand\eea{\end{eqnarray}}
\newcommand\bea{\begin{eqnarray}}

\newcommand\TeV{\,\mbox{TeV}}
\newcommand\GeV{\,\mbox{GeV}}
\newcommand\MeV{\,\mbox{MeV}}
\newcommand\keV{\,\mbox{keV}}

\newcommand\mpl{M_{\rm P}}

\newcommand\lsim{\mathrel{\rlap{\lower4pt\hbox{\hskip1pt$\sim$}}
    \raise1pt\hbox{$<$}}}
\newcommand\gsim{\mathrel{\rlap{\lower4pt\hbox{\hskip1pt$\sim$}}
    \raise1pt\hbox{$>$}}}

\newcommand\diff{\mbox d}

\def\dslash{\not{\hbox{\kern-2pt $\partial$}}}
\def\Dslash{\not{\hbox{\kern-4pt $D$}}}
\def\Oslash{\not{\hbox{\kern-4pt $O$}}}
\def\Qslash{\not{\hbox{\kern-4pt $Q$}}}
\def\pslash{\not{\hbox{\kern-2.3pt $p$}}}
\def\kslash{\not{\hbox{\kern-2.3pt $k$}}}
\def\qslash{\not{\hbox{\kern-2.3pt $q$}}}

 \newtoks\slashfraction
 \slashfraction={.13}
 \def\slash#1{\setbox0\hbox{$ #1 $}
 \setbox0\hbox to \the\slashfraction\wd0{\hss \box0}/\box0 }
 
\def\ee{\end{equation}}
\def\be{\begin{equation}}

\newcommand\sub[1]{_{\rm #1}}

\newcommand\mgrav{m_{3/2}(t)}
\newcommand\mgravsq{m_{3/2}^2(t)}
\newcommand\mgravvac{m_{3/2}}

\begin{document}

\begin{flushright}
LANCS-TH/9924  
\\hep-ph/9912313\\
(December 1999)
\end{flushright}
\begin{center}
{\Large \bf Late-time creation of gravitinos from the vacuum}

\vspace{.3in}
{\large\bf  David H.~Lyth}

\vspace{.4 cm}
{\em Department of Physics,\\
Lancaster University,\\
Lancaster LA1 4YB.~~~U.~K.}

\vspace{.4cm}
\end{center}

\begin{abstract}
Starting with the vacuum fluctuation, it is  known that gravitinos will be 
created just after inflation, with number density  $\sim 10^{-2}M^3$ where 
$M$ is the mass of the inflaton. Here, we argue that creation may be expected
 to continue, maintaining about the same number density,  until a usually 
much later epoch. This epoch  is either  the `intermediate epoch' when Hubble
parameter falls below the gravitino mass, or the reheat epoch if that is 
earlier. We verify that such late-time creation indeed occurs if  only a 
single chiral superfield is relevant, using the description of the helicity 
$1/2$ gravitino provided recently by   Kallosh et. al. (hep-th/9907124) and 
Giudice et. al. (hep-ph/9907510). Arguments are presented in favor of 
late-time creation in the general case. For the usual inflation models,
$M$ is rather large and  gravitinos from late-time creation are  so abundant 
that a subsequent era of thermal inflation is needed to dilute them.
\end{abstract}

\paragraph{Introduction} 
Gravitinos are created in the early Universe by 
thermal collisions \cite{subir},  and some time ago
 \cite{lrs}
it was pointed out 
that they will  also be created non-thermally, starting  from 
the vacuum fluctuation that exists  well before horizon exit during
 inflation. To see whether creation from the vacuum is significant,
one needs  equations which describe the evolution of the
 gravitino mode functions. Such equations  have recently been presented
 \cite{kklv,grt}, for the case that 
only one chiral superfield is relevant.
 Using them, the 
 authors find that  the number density  of gravitinos created just after
inflation is  of order $10^{-2}M^3$, where $M$ is the mass of the inflaton
after inflation, and they have noted that these gravitinos may be more 
abundant after  reheating than those created from thermal collisions.
 A  similar result has
been obtained for hybrid inflation, which involves  at least two 
fields,  assuming  that in this context the gravitino becomes
the goldstino of global supersymmetry \cite{grt2} (see also \cite{mp}).

In this note, we argue that  this is unlikely
to be  the end of the story \cite{p99new}. Rather, creation  is likely to
continue, maintaining
about the same number density,  until 
 either  the `intermediate epoch' when Hubble
parameter falls below the gravitino mass, or the reheat epoch if that occurs 
earlier. 
We begin by verifying that such late-time creation indeed occurs
if  only a single chiral
superfield is relevant, using the
the description of the helicity $1/2$ gravitino provided recently
in \cite{kklv,grt}.
Then we consider the case where other fields are relevant;
a different field to break supersymmetry in the vacuum, a third
field to allow hybrid inflation, and the fields  corresponding to
particles  created just after inflation
by  preheating. In all these cases, we argue that late-time creation will
continue until the intermediate epoch, unless it is terminated by
preheating. Finally, we consider the  cosmological
consequences of late-time creation, which are rather dramatic
in the most popular models of inflation.

\paragraph{Describing the gravitino in the early Universe}
To   calculate the abundance of gravitinos created from the vacuum,
one  needs equations describing the evolution of the mode functions
for the helicity  $1/2$ and $3/2$  gravitino states, as seen by
a comoving observer in the expanding Universe. 

For  the helicity $3/2$ state, one can safely proceed 
by using the 
Rarita-Schwinger equation
with appropriate constraints, evaluated in  Robertson-Walker
spacetime, and with time-dependent
 mass $\mgrav$ given by the usual
$N=1$ supergravity formula. (We denote the  vacuum value of $\mgrav$
by $\mgravvac$ without an argument.)  A suitably-defined 
mode function satisfies the spin $1/2$ equation, 
 with  the same effective mass $\mgrav$ that appears in the
Rarita-Schwinger equation
\cite{mm,kklv,grt,l}. As $\mgrav$ can hardly be bigger than
the Hubble parameter $H$ if it is to vary non-adiabatically,\footnote
{The $N=1$ supergravity expression for $H^2$ contains a term
$-\mgravsq$, which can hardly vary rapidly if it is   canceled to
high accuracy  by the rest of the expression. Of course, a precise
cancellation does occur at the present time, for some unknown
reason (the cosmological constant problem).}
 the abundance of  helicity $3/2$ gravitinos created from the vacuum
will no bigger than that of modulini created from the same
mechanism, which is negligible \cite{lrs,nont} compared with  the abundance
of gravitinos from thermal collisions.

 For the helicity $1/2$ state, the same procedure may in general require
modification, because the would-be goldstino (existing in the limit
of global supersymmetry) may be a time-dependent mixture of
the  spin $1/2$ fields. To avoid this problem, the case 
where only a single chiral supermultiplet is relevant has been studied
\cite{kklv,grt}.
It is found     that a  suitably-defined mode function again satisfies 
the spin $1/2$ equation, but with a different effective
mass $\tilde m(t)$. 

In a general model of inflation, the idealization of a single chiral
supermultiplet will not be adequate. In hybrid inflation 
models one needs, in addition to the slowly rolling inflaton field,
one or more additional fields to provide the constant part of the potential.
Even if the inflaton field is the only one relevant for inflation,
a second field is generally needed to break supersymmetry in the vacuum.
Even so,  the possibility that the inflaton field is the only relevant
one is not excluded.
Also,  the field  responsible for supersymmetry breaking may 
at some stage oscillate and dominate the energy density of the Universe,
even if it is not responsible for inflation. Let us proceed on the assumption
that only a single field is involved in gravitino creation, so that 
the equations presented in \cite{kklv,grt} can be used.

\paragraph{Gravitino creation from the oscillation of the 
supersymmetry-breaking
field}
The 
 effective mass $\tilde m(t)$ of the helicity $1/2$ mode function 
is given by \cite{kklv,grt}\footnote
{In  Eq.~(3.23) of  \cite{kklv}, the signs of the second and third terms
should be reversed \cite{renata}, after which  that equation becomes
identical with \eq{mtil} upon making the identification $a^{-1}\Omega\sub L
\equiv -\tilde m(t)$. 
The sign of the effective mass is not physically significant, and in 
particular it will not affect the gravitino abundance. (Equivalently,
the sign of the last term of \eq{mode} is not physically significant,
since it corresponds to replacing a mode function by its complex conjugate.)
We have checked  \cite{nont} that \eq{mtil} is equivalent to the more
complicated expressions given in \cite{grt}.
In both \cite{kklv} and \cite{grt},
it is  assumed that the field is canonically normalized, but
the results given there  are valid also for arbitrary normalization
\cite{personal}. In fact, for 
 a single real scalar field, one  can  always transform
at least locally 
from arbitrary normalization to canonical normalization.
 In the supergravity model, $\phi$ is the real part of a complex field,
and the corresponding  transformation of the complex field
is at least locally holomorphic, leading to an equivalent supergravity
theory.}
\be
\tilde m(t)= \mgrav -\frac32 \mgrav (1+A_1) -\frac32 H A_2 - \mu 
\,,
\label{mtil}
\ee
where
\bea
A_1 &\equiv& \frac{P-3\mpl^2 \mgravsq}{\rho+3\mpl^2\mgravsq} 
\label{A1} \\
A_2&\equiv &\frac23\frac{3\mpl\dot \mgrav}{\rho+3\mpl^2\mgravsq} 
 \label{A2}\\
A&\equiv& A_1 + i A_2 = e^{i\chi} \label{A}\\
\mu&\equiv & \frac12 \dot\chi \label{mu}
\,.
\eea
In this expression, $\mpl=2.4\times 10^{18}\GeV$ is the Planck scale.

The   energy density $\rho$ and the pressure $P$ are supposed to be  dominated
by a real scalar field $\phi$. Taking its kinetic term to be
canonical, 
\bea
\rho &=& V +\frac12\dot\phi^2 \label{rho}\\
P &=& -V +\frac12\dot\phi^2 \label{P}
\,,
\eea
where $V(\phi)$ is the potential.
The continuity equation is 
$\dot \rho=-3H(\rho+P)$, equivalent to the field equation
\be
\ddot \phi + 3H\dot\phi + V' =0
\label{feq}
\,
\ee
where $H$ is the Hubble parameter, related to the energy density by
$3\mpl^2 H^2=\rho$. 

For future reference, let us note that the equations for
$A_1$, $A_2$ and  $\mu$
may be written in terms of $H$, $w\equiv P/\rho$, and $\mgrav$,
with the time-derivative of this last quantity eliminated
using \eq{A2}. One finds
\bea
A_1 &\equiv& \frac{w H^2-\mgravsq}{H^2+\mgravsq} \label{A12}\\
A_2&\equiv & \frac{\[1-w^2 + 2\(1+w\) \mgravsq/H^2 \]
^\frac12}{1+\mgravsq/H^2} \\
\mu &\equiv & \frac32 \mgrav(1+A_1) -\frac12\frac{\dot w - 3H(1+w)(w-A_1)}
{A_2(1+\mgravsq/H^2)} \label{mtil2} 
\,.
\eea

For momentum $k/a$, a
 suitably defined helicity $1/2$  mode function satisfies the equation
\be
u'' + \( k^2 + (a\tilde m)^2 + i (a\tilde m)' \) u = 0
\label{mode}
\,,
\ee
where  the prime denotes differentiation with respect to conformal time
$\diff \eta=\diff t/a$, and $a$ is the scale factor of the Universe
such that $H=\dot a/a$.
To calculate the gravitino abundance, one starts at early times
with the negative-frequency solution $u\propto e^{-i\omega \eta}$,
corresponding to the vacuum.
At late times there is  a linear combination of positive
and negative frequency modes, and the occupation number
is the coefficient of the positive frequency mode.
Significant production occurs with 
momentum $k/a$ if 
there is appreciable violation of a weak adiabaticity condition
\cite{nont}
\be
|\overline{(a\tilde m)'}| \ll \omega^2\equiv k^2 + (a\tilde m)^2
\label{wad}
\,,
\ee
where the 
average is over a conformal time interval $\omega^{-1}$. 
 In practice $k\sub{max}$,
 the biggest $k$ for which significant creation occurs,
is simply the biggest value achieved by $a\tilde
m$, within the regime
where $\tilde m$ varies non-adiabatically ($|\dot {\tilde m}
|\gsim \tilde m^2$).

Let us follow the evolution of $\tilde m$, to estimate $k\sub{max}$.
During inflation, $\tilde m\sim \mgrav$ is much less than $H$ in 
magnitude. After inflation,  $\phi$ oscillates about its vacuum
value which we set equal to zero so that
$\phi\simeq \phi_0(t)\sin M t$ where $M\gg H$ is the  inflaton mass.
We now have
  $\tilde m\simeq -\mu
\equiv -\frac12\dot\chi$, with
\bea
\cos\chi &\equiv& \frac{\frac12\dot\phi^2 -V -3\mpl^2\mgravsq}
{\frac12\dot\phi^2 +V+3\mpl^2\mgravsq} \\
&\simeq& \frac{\cos 2Mt -\mgravsq/H^2}
{1+\mgravsq/H^2} 
\,.
\label{coschi}
\eea
When $\dot\phi=0$, $\cos\chi=-1$. The crucial point is that the 
maximum value of $\cos\chi$ is {\em less} than $+1$, because
$\cos\chi=1$ would correspond to  $V=-3\mpl^2\mgravsq<0$.
It follows that 
 $\chi$ oscillates   in some  range $\epsilon < \chi
< 2\pi-\epsilon$. With the reasonable assumption $\mgravsq
\sim \mgravvac^2$, $\epsilon \sim \mgravvac/H\ll 1$.
While $\chi$ is rising, $\tilde m\simeq -M$, and while
it is falling, $\tilde m\simeq +M$.
Because of the sign switches, the evolution of the mode function
is non-adiabatic, and gravitino creation continues, with every-increasing
comoving momentum $k\sim aM$.

When $H$ becomes of order $\mgravvac$, the sudden switches in sign
of $\tilde m$ give way to a complicated variation, still on the timescale
$M^{-1}$ and with typical value of order $M$. (If desired, this
variation is conveniently calculated from \eqst{A12}{mtil2},
with $w=\cos 2Mt$.)
 Creation finally ceases
only when $H$ falls  below $\mgravvac$, and according to \eqs{mtil}{mtil2}
$\tilde m$ falls smoothly to the true value $\mgravvac$, recovering
the flat spacetime description of the gravitino.

We studied this example primarily to  provide an existence
proof, that late-time gravitino creation occurs in at least one case. This is 
 the  case that the oscillating field responsible for the 
energy density of the Universe, is the same as the field
breaking supersymmetry in the vacuum.
 Before moving on, we note that this case may  by realized in Nature
if supersymmetry breaking is gravity mediated, though the mass
$M$ would then 
 probably be too small for late-time creation to be significant.
Indeed,  superstring theory
suggests the existence of several fields (moduli) with only gravitational
strength interactions, and mass very roughly of order $\mgravvac$.
(K\"ahler stabilization of the moduli can give  masses several
orders of magnitude bigger, for instance a dilaton mass of order $10^6\GeV$
is found in  \cite{glm}.) 
At least some of the  moduli 
may well be displaced from the vacuum at the end of inflation,
and one of them might be  the inflaton). Any displaced modulus
 at first tracks the changing vacuum value, but it  starts to oscillate
at the intermediate epoch  \cite{thermal2},
and then   dominates
the energy density. Finally, one of the moduli is usually supposed
to be responsible for supersymmetry breaking
in the gravity-mediated case. The case that we have studied will be
realized if this modulus is also the inflaton, and is the only
one that oscillates.
As far as the oscillation regime is concerned, the case  is also
realized  if 
the inflaton is different from the supersymmetry breaking modulus,
but  decays early so that the latter is the only oscillating field
at  the intermediate epoch.

\paragraph{Late-time gravitino creation in the general case}
When additional fields and/or particles are involved, 
equations describing the evolution of the gravitino field
are not yet available. However, late-time creation seems  likely to occur
quite generally up to the intermediate era, unless reheating occurs
first. 

Consider first models where the slowly-rolling inflaton field
is the only one relevant for inflation, but something 
else breaks supersymmetry  in the vacuum while ensuring that 
the potential vanishes there. In the model where the inflaton 
field did both jobs, 
late-time creation occurred  because
supergravity corrections to 
global supersymmetry become very important every time the potential
the potential dips to zero, causing the mode function to vary
non-adiabatically. One should not expect that the introduction of
something else to break supersymmetry, would restore the
adiabaticity.

Next, consider hybrid inflation models, where two fields are
 oscillating after inflation, with presumably a third breaking supersymmetry
in the vacuum. In  supersymmetric models, the two oscillating
fields typically have the same mass $M$ after inflation.
 In general they are not oscillating
in phase (though see \cite{mar} for a case where they are),
 which means that they cannot be replaced by a single field.
Just after inflation, gravitino creation in this situation may
be estimated  in the global supersymmetry limit, as described
in \cite{grt2} for the case of $F$-term models.
 The number density will again be of order $M^3$,
since the inflaton mass climbs to that value in a time of order
$M^{-1}$. Again, there is no reason to   expect gravitino production
to stop, since supergravity corrections will again become important 
every time the potential dips to zero.

In both of these situations, a 
definite calculation   will clearly become possible
when the supergravity formalism  of
\cite{kklv,grt} is  extended to include two or more chiral
superfields. It is not so clear how to calculate things
 when particles
as opposed to homogeneous fields become important, but 
still it is fairly clear what will happen.
Consider first the case that 
 preheating \cite{preheat} converts most of the 
oscillating field energy into marginally relativistic particles.
The energy of the latter will be somewhat reduced by redshifting
even if it does not decay  promptly into radiation,
 and one expects
that after a few Hubble times  oscillating fields  again
 account for a non-negligible
proportion of the energy density. Then  
 late-time gravitino creation will presumably continue  until the intermediate
era, unless reheating happens first. After reheating, practically
all of the energy density is in radiation, and one expects
creation to stop, because nothing is varying rapidly.
We emphasize that, although these expectations look quite reasonable,
it is at the moment totally  unclear how to describe the gravitino,
when  supersymmetry breaking comes mainly from the particle gas
in the early Universe.

\paragraph{Cosmological significance of late-time gravitino production}
After production stops,
the   occupation number will be of order 1 below
 $k=k\sub{max}$,
 giving number density \cite{lrs}
\bea
n &\simeq& \frac2{4\pi^2} a^{-3} \int^{k\sub{max}}_0 k^2 dk \\
&\sim &10^{-2} (k\sub{max}/a)^3 
\,.
\eea
The number density when creation stops is $n\sim 10^{-2} p^3$, where
$p=k\sub{max}/a$ is the maximum momentum of the created gravitinos.

We focus on the extreme case, where    gravitino creation
ends only at the intermediate era.
This corresponds to energy density of order $M\sub S^4$,
where $M\sub S \simeq \sqrt{\mpl\mgravvac}$ is the scale
of supersymmetry breaking.
 In gravity-mediated models of supersymmetry 
breaking, $\mgravvac\sim 100\GeV$ and $M\sub S\sim 10^{10}\GeV$.
In typical gauge-mediated models, $\mgravvac\sim 100\keV$ 
and $M\sub S\sim 10^7\GeV$.

At the intermediate era, the gravitino number density is of order
$10^{-2}M^3$. 
The relative abundance at 
nucleosynthesis is  therefore \cite{lrs}
\bea
\frac ns 
&\simeq& 10^{-2}\frac {\gamma T\sub R M^3}{M\sub S^4}\\
&\simeq& 10^{-2}\frac {\gamma T\sub R M^3}{\mpl^2\mgravvac^2}
\,.
\label{nvac}
\eea
Here,  
 $s$ is the entropy density at nucleosynthesis,
and $\gamma^{-1}$ is the increase in entropy per comoving volume
(if any), between reheating at
temperature $T\sub R$ and nucleosynthesis. 
If the entropy increase
 comes only from a late-decaying particle, there can be only
a modest increase corresponding to  $\gamma\sim T\sub{FR}/T\sub{EQ}$, where
EQ is the  epoch when the particle first
dominates the energy density, and FR is the final reheat
epoch when it decays. This gives  $\gamma T\gsim T\sub{FR}\gsim
10\MeV$, where the bound comes from nucleosynthesis. However,
$N$ $e$-folds of 
thermal inflation \cite{thermal1,thermal2,thermal3} 
 may also occur. This would  reduce
$\gamma$ by an additional
 factor $e^{-3N}$. One bout of thermal inflation
typically gives  \cite{thermal2,thermal3}
$N\simeq 10$  and a total $\gamma$ perhaps of order
$10^{-15}$. Two bouts might give $N\simeq 20$ and up to 
$\gamma\sim 10^{-30}$ \cite{thermal2}.

The cosmological significance \cite{subir} of the gravitino depends on its 
true mass $\mgravvac$. With gravity-mediated  supersymmetry breaking,
one expects  $\mgravvac\sim 100\GeV$ to $1\TeV$,
and observation then requires 
\be
n/s\lsim 10^{-13}
\,. 
\label{nsyn}
\ee
The abundance of gravitinos from thermal 
collisions is then  
\be
n/s\sim 10^{-13} (\gamma T\sub R/10^9\GeV)
\,,
\label{1}
\ee
 leading to
the bound $\gamma T\sub R\lsim
10^9\GeV$.\footnote
{We refer here to the  thermal creation at the initial reheating.
If thermal inflation subsequently occurs,  the most significant thermal
creation occurs at the final reheating, and $\gamma T\sub R$
in \eq{1} is to be
replaced by the final reheat temperature.}
 Using instead \eq{nvac}, we find
\be
10^{13}
\frac ns \sim \(\frac{M}{10^{7}\GeV} \)^3\(\frac{ \gamma T\sub R}
{10^9\GeV} \) \( \frac{100\GeV}{\mgravvac} \)^2\lsim 1
\label{22}
\,.
\ee
Gravitinos created from the 
vacuum are
more abundant than those from thermal collisions, if $M\gsim
10^7\GeV$. Without thermal inflation, we need $\gamma T\sub R \gsim
10\MeV$ and therefore $M\lsim 10^{11}\GeV$. In the worst case
$T\sub R\sim M\sub S$, even a single bout of thermal
inflation allows only $M\lsim 10^{11}\GeV$. With two bouts, or with
one bout and low $T\sub R$, more or less any $M$ up to $\mpl$ might
be accommodated.

With  gauge-mediated  supersymmetry breaking, one expects
very roughly  $1\keV \lsim \mgravvac\lsim 100\GeV$, with the upper
decades disfavoured. Unless $\mgravvac\gsim 100\MeV$, this leads
to  a stable gravitino
with  present density 
\be
\Omega_{3/2} \simeq 10^5 \( \frac{\mgravvac}{100\keV } \)
\frac ns \lsim 1
\ee
If creation from thermal collisions dominates, 
then very roughly \cite{subir}
\be
\Omega_{3/2} \sim\(\frac{100\keV}{\mgravvac} \) \(\frac
{\gamma T\sub R}{10^4\GeV} \)
\,.
\ee
Using instead
\eq{nvac}, 
\be
\Omega_{3/2} \sim \(\frac{M}{10^8\GeV} \)^3
\(\frac{100\keV}{\mgravvac} \) \(\frac
{\gamma T\sub R}{10^4\GeV} \)  \lsim 1
\label{25} 
\,.
\ee
In this case, creation from the vacuum is more efficient if
$M\gsim 10^8\GeV$, and we need  thermal inflation if  $M\lsim
10^{10}\GeV$. In the favored case
$\mgravvac\sim 100\keV$, corresponding to $M\sub S\sim
10^7\GeV$, the constraints on $M$ are similar to those
in the gravity-mediated case. 

So much for the  case that reheating happens
after the intermediate epoch corresponding to $T\sub R\lsim
M\sub S$. By way of contrast, consider the opposite extreme
of instant reheating, $T\sub R\sim V^{1/4}$.
Then,
\be
\frac n s \sim 10^{-2} \gamma \(\frac M{V^\frac14} \) ^3
\,.
\label{ns2}
\ee
Either $M/V^{1/4}$ should be sufficiently small, or 
 thermal inflation is again needed.

\paragraph{Conclusion}
Let us end by considering the implications of late-time gravitino
creation for models of the early Universe, and in particular of
inflation.
 Supersymmetric models of inflation are reviewed for example
in \cite{treview}.
In some of them, notably those using only flat
directions, the mass of the field(s) oscillating after inflation
is small, and late-time creation cannot be significant.
However, in the most  models, normalized to give the correct prediction for
large scale structure, 
the inflaton mass is bigger than
$10^{10}\GeV$.  Using the popular names, examples 
 include chaotic inflation, 
the most popular hybrid inflation models ($D$ term,
and  the usual $F$ term model)
and most  new inflation potentials. The popular  hybrid inflation 
models require $M\sim V^{1/4}\sim 10^{15}\GeV$ or so.

In all of these cases, the possibility of late-time gravitino creation
 drastically
changes ones view about the reheat temperature.
Starting the discussion with instant reheat, \eq{ns2} represents
a significant constraint. The constraint becomes {\em stronger}
as the reheat temperature is lowered,
because gravitino creation persists to a later epoch so that
$n/s$ increases. Only when the reheat temperature falls below
 $M\sub S$, does the constraint, now represented by
   \eqs{22}{25}, start to become weaker. 	In many
cases, the constraint cannot be met simply by lowering the reheat
temperature. Instead, the entropy dilution factor $\gamma$ has
to be small, often so small that thermal inflation is required.

To avoid these constraints, one might turn to models using
only flat directions, leading to $M$ perhaps of order $\mgravvac$,
and giving negligible gravitino creation. Such models include
the modular inflation mentioned earlier (so far without a concrete
realization in the context of string theory) and certain hybrid inflation
models.

The next step will be to check that late-time gravitino production
actually occurs in the models mentioned, using equations
that describe the gravitino in the presence of several fields.
After that would come the more challenging task of describing the 
gravitino when a particle gas is the dominant source of supersymmetry
breaking.

\subsubsection*{Acknowledgments}

I am indebted to Andrei Linde, Renata Kallosh and Toni Riotto for useful
discussions.

\newcommand\pl[3]{Phys. Lett. #1 (19#3) #2}
\newcommand\np[3]{Nucl. Phys. #1 (19#3) #2}
\newcommand\pr[3]{Phys. Rep. #1 (19#3) #2}
\newcommand\prl[3]{Phys. Rev. Lett. #1 (19#3) #2}
\newcommand\prd[3]{Phys. Rev. D #1 (19#3) #2}
\newcommand\ptp[3]{Prog. Theor. Phys. #1 (19#3) #2}
\newcommand\rpp[3]{Rep. on Prog. in Phys. #1 (19#3) #2}
\newcommand\jhep[2]{JHEP #1 (19#2)}
\newcommand\grg[3]{Gen. Rel. Grav. #1 (19#3) #2}

\end{document}